# Fabrication of a Low-Cost Real-Time Mobile ECG System for Health Monitoring

Soheil Khooyooz[1,2], Mohammad Ali Ahmadi Pajouh[1*], Kamyab Azizi[3]

[1]Department of Biomedical Engineering, Amirkabir University of Technology (Tehran Polytechnic), Tehran, Iran

[2]Department of Electrical Engineering, Sharif University of Technology, Tehran, Iran

[3]Department of Electrical Engineering, Amirkabir University of Technology (Tehran Polytechnic), Tehran, Iran

[*]Corresponding Author

## Abstract

The electrocardiogram (ECG) signal is one of the most vital signals that can be used to investigate the performance of heart. Based on the ECG graph, we can identify different disorders and diseases. Therefore, monitoring this signal is of great importance. Many electrodes (usually 12) are employed to acquire this signal in clinics and hospitals; therefore, a nurse must install them on the body of the patient to record the signal. In this project, we built a device that can acquire the real-time ECG signal and display it on the mobile screen with the least number of electrodes (three electrodes) without requiring a nurse to install the electrodes using the simplest and cheapest type of ICs and transmitters including STM32F030F4P6 microcontroller, AD620 instrumentation amplifier, TL084 amplifier, TC7660 voltage converter, LM1117 regulator, and HC-05 Bluetooth module. Moreover, the device was designed in a way that could operate using a single 9V battery or power adaptor. First, to amplify the signal and remove a part of the its noise, the ECG signal is given to the primary analog circuit. Then, in the digital section, using a microcontroller, the signal is discretized, processed, and finally transmitted to a mobile phone for final processing, information extraction, and displaying. In the mobile-written application, we developed a mathematical algorithm based on Pan-Tompkins algorithm to calculate the heartbeat rate, in which the signal peaks are determined after several processing steps using a threshold. In this regard, the ECG signals of 10 subjects were recorded and analyzed to calculate the optimum threshold. Finally, the power spectral density (PSD) and input referred noise were calculated and plotted to check the output signal quality. Based on the PSD and input referred noise amplitudes, we achieved a signal-to-noise ratio of 50dB.

## 1 Introduction

The design and expansion of biosensor systems with the purpose of health monitoring has attracted a lot of attention in the scientific community and industry in recent years [1]. Due to the increase in health costs and the aging of the global population, the need to monitor the patient's health status in their home and in situations where they do not have access to a hospital is intensely felt. Real-time monitoring devices report the patient's health status to the patient, a medical center, or a specialist and can alarm in particular situations. Moreover, these systems can be employed to manage and monitor the different statuses of chronic diseases, older people, patients needing rehabilitation, and people with particular disabilities. The primary motivation for using such systems is high health and maintenance costs, the advancement of biosensor display technology, intelligent textiles, microelectronics, and wireless communication. These systems can consist of several different types of small physiological sensors, transmitter modules, and processing units, which can be used for health monitoring throughout the day in any place, mental state, and activity. Simultaneous multi-parametric recording of physiological signals and analysis of the relationship between them can be helpful in the early identification of some diseases. This project aims to record the ECG signal, calculate the number of heartbeats, and display it on the mobile screen. For the patient to do so conveniently, the least number of electrodes are used to record the signal, and a Bluetooth module is employed to transmit information to the mobile phone [1]. The ECG signal is one of the most critical biological signals, and recording and displaying it is vital; because it shows how the heart muscle works, and if there is an arrhythmia in the heart's function, it can be discovered from this signal. Hence, according to the problems and the necessary preparations attributed to the ECG signal acquisition methods in hospitals and clinics, as well as some patients' need for checking their heart's signal continuously, there is a great need for a more accessible device that can display the real-time signal and make the acquisition process more straightforward.

Much research has been conducted in the field of ECG signal acquisition. This part reviews the projects related to ECG signal acquisition and its wireless transmission to distant centers. In 2001, Oleg Orlovo et al., recorded the ECG signal with a single lead and transmitted it to a medical center via telephone by converting it into an audio signal. Many subjects participated in the performance evaluation stage of this project. During the examinations, the subjects in their homes attached electrodes to their body, sending their ECG signals via telephone cables to medical centers. Finally, a doctor in the center could observe the real-time form of the subjects' ECG signals [2]. Samuel de Lucena and Daniel Sampaio, 2015, built an ECG signal recording device in which the signal data was transmitted serially to a mobile phone via Bluetooth. In this project, they developed a mobile phone application using Android Studio software [3]. Ying Bai and Robert Matthews, 2006, built a device to record the ECG signal wirelessly; for ease of utility, they employed capacitive isolation sensors instead of conventional electrodes to record the signal [4]. In conventional electrodes, the presence of the gel, contact with the surface of the hand, their separation in case of movement, and sweating of the skin, reduce the tendency to use them in recording vital signals. The recorded signal using the aforementioned capacitive sensors and conventional electrodes had a 99% correlation. Therefore, in their project, two capacitive sensors were installed on the back of the subject's shirt (in which the electrodes stood on the chest after wearing the shirt) to record the ECG signal. After the signal was recorded, it was sent to a laptop via transmitters attached to the electrodes [4]. Ishani

Mishra and Sanjay Jain, 2022, developed a compressive sensing framework to enhance the ECG signal reconstruction quality while transmitting it wirelessly using Rain Optimization Algorithm (ROA) and Sparsity Adaptive Matching Pursuit Algorithm (SAMPA) [5]. R. V. Kapse and A. S. Barhatte, 2022, fabricated an ECG signal acquisition device for analyzing normal and abnormal signals using AD8232 sensor to extract, amplify, and filter the ECG signal. Then, they employed an HC-05 Bluetooth and a Wi-Fi module to transmit the signal to an android application in cell phone which was developed using MIT app inventor [6]. G. V. Hueh et al., 2022, developed an integrated in-house ECG monitoring device for detection and classification of different cardiovascular states including bradycardia, tachycardia, and normal sinus rhythm. They transmitted the signal to an android application which was developed in Android Studio via a Bluetooth device. The recorded signals in mobile phones were converted to text files and stored in cell phone local memory. They also have implemented a Firebase Authentication and Firebase Storage based on Google Cloud for data management which allows patients and caretakers secure access to data in the future using Google Cloud platforms [7]. T. Dubey et al., 2022, fabricated a device to acquire ECG signal, body temperature, and blood oxygen level to monitor vital parameters of body in real time. They sent these parameters to an android application developed in Android Studio wirelessly using a Bluetooth module. They employed AD8232 to record the ECG signal [8]. D. F. H. Sadok, 2023, developed a low-cost real-time wireless ECG device for monitoring the health of newborns. They employed an ultra-low power wireless transceiver with embedded base-band protocols and a BLE communication module to transmit the ECG signals to an android application in mobile phone [9]. Suhiam Sarbaras J.S., 2022, fabricated an ECG signal system for long-time heart monitoring. They employed HC-05 Bluetooth module to transmit the signal wirelessly to an android application developed in MIT app inventor. Then, they stored the recorded signal in cell phone SD card for later analysis of the signal [10].

The rest of this paper is organized as follows. In Section II, the designation of the hardware and software parts of the device, along with their implementation, will be deeply investigated. In section III, our results and findings are discussed. Finally, the conclusions are given in section IV.

## 2 Methodology

In this section, the required hardware and the circuit designation along with the details and tips related to each part, will be investigated.

### 2-1 Schematic designation

The amplitude of the ECG signal is about 1 to 10mV; hence, circuits with a high gain and accuracy must be designed to measure this signal. We employ a microcontroller and a Bluetooth to transmit the recorded signals to the mobile phone. These digital modules can inject high-frequency noise into the circuit and cause measurement errors due to the signal's low amplitude. We must isolate the analog circuit from the digital circuit using a ferrite bead to avoid this problem. Ferrite beads prevent the leakage of high-frequency signals from the digital circuit to the analog circuit.

### 2-1-1 STM32F030F4P6 microcontroller

The simplest and cheapest type of STM32 microcontroller, which has a flash memory of 16 KB and a RAM memory of 4 KB, can meet the requirements of this project [11]. After recording the ECG signal and passing it through the primary filters, the signal is biased to direct current (DC) voltage and given to the input of the analog-to-digital converter (ADC) of the microcontroller. Then, the data is processed in the microcontroller and transmitted to the phone through the universal synchronous/asynchronous receiver/transmitter (USART).

### 2-1-2 Power supply

The circuit's power supply must feed amplifiers, the microcontroller, and the Bluetooth. The amplifiers used in this project are TL084 and AD620 ICs; These ICs need both a positive and a negative power supply for a correct operation. On the other hand, this project aims to design a device with little complexity to make the ECG signal recording process easier. Therefore, the device is designed with only one battery, and the negative supply is provided with the help of TC7660 IC for the amplifiers. According to the figure of the output voltage versus output current in the IC's catalog; the IC's output voltage is like an ideal voltage source connected in series with a 70Ω resistor [12]. According to this figure, if the output current is around 20mA, The IC's output voltage will be about -7/6V (in practice, this current is about 10mA). This amount of voltage is sufficient to drive other ICs in the circuit. The power required by the Bluetooth and microcontroller is 3.3V, created with the help of a LM1117 voltage regulator. The device's current consumption is about 70mA, most of which is related to the transmitter; using a 200mAh 9V battery, we can use the device for 3 hours.

### 2-1-3 Circuit's input stage

The ECG signal has a differential nature, and its amplitude is very low; hence, we need a differential amplifier with a high common mode rejection ratio (CMRR) for the input stage. AD620 can meet this requirement; it is an instrumentation amplifier used in medical applications such as recording ECG and blood pressure signals due to its unique features [13]. The input stage is designed as Figure 1.

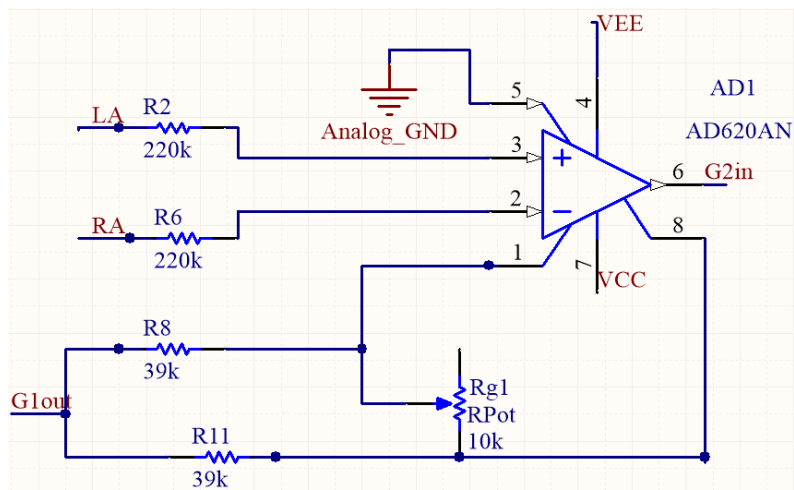

*Figure 1 The analog circuit input stage*

### 2-1-4 Right leg driver (RLD)

Electromagnetic waves, power line interference, noise from other devices, and high-frequency noise can all interfere with the ECG signal. This interference takes place both through the device and the human body. The human body acts like an antenna and receives and amplifies some of these interference waves directly. The voltage formed in this way has a significant amplitude, and if it is not eliminated, it's practically impossible to record the ECG signal. A circuit with negative feedback is needed to remove the aforementioned interfering signals. This circuit will be connected to the right leg and is called the right leg driver [13] (RLD). The RLD circuit is illustrated in Figure 2.

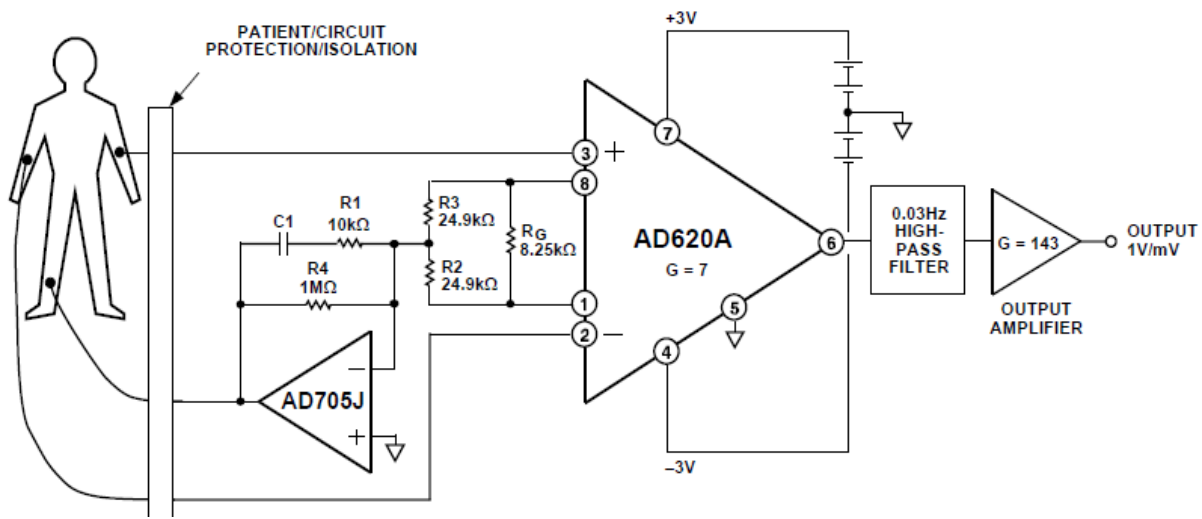

*Figure 2 The schematic of Right Leg Driver circuit [10]*

## 2-1-5 High-pass, low-pass and notch filters

The frequency range and amplitude of normal people's ECG signals range from 0.05 to 100 Hz and 1 to 10mV, respectively [14]. The electromagnetic waves in our surroundings enter the ECG circuit through the cables of the measuring device and affect the signal. If they are not filtered, errors will occur during measurements [15]. Among these noises, removing power line noise whose frequency is 50Hz is of more importance because it is in the frequency range of the ECG signal and can affect its components. A notch filter with the unity gain is employed to filter the power line noise with a frequency of 50Hz. The highest-frequency component in the ECG signal, the R wave, has a frequency of about 100Hz; therefore, a low-pass filter with a cutoff frequency of 105Hz and a gain of 11 is employed to remove noises having frequencies higher than 100Hz. Another problem regarding ECG signal recordings is a deviation from the main line of the signal, known as the baseline wander. Baseline wander or low-frequency noise, can be attributed to factors such as breathing, coughing, the movement of the right leg, temperature changes, and the presence of bias in the instrumentation amplifier. Its frequency range is 0.05 to 2Hz [15]; to eliminate the low-frequency noise, we employed a high-pass filter with a cutoff frequency of 0.1 Hz and a gain of 8. Note that the resistors R6, R10, R16, and R19 are used to adjust the gain value of the filters. According to each stage's gain and the unity gain of the last stage, the total gain of the circuit is 500. The ECG signal amplitude is about 1mV in the Lead I and about 5mV in the Lead II; hence, this gain is sufficient to amplify the source. According to the maximum possible swing of the circuit, which is 3V, the signal saturates for the gain value of more than 500. The circuit and graph of amplitude and phase response of the corresponding filters are illustrated in Figures 3.

For designing the abovementioned filters, we used TL084 amplifier IC which has four internal amplifiers with a high input impedance and low input bias and offset current making it suitable for our purpose [16].

## 2-1-6 Biasing the last stage output signal

Considering that the ADC input of the microcontroller must be positive, the output signal of the last stage of the analog circuit must be biased to a positive DC voltage. The voltage of the signal that enters the ADC in the employed microcontroller must be from 0 to 3.3V. As shown in Figure 4, we used a voltage divider to bias the signal.

According to the resistances' values and the 3.3V power supply, the signal is biased to a 0.3V voltage as follows:

$$Bias\_out = 3.3 \times \frac{R22}{R21 + R22} = 0.3V \qquad (1)$$

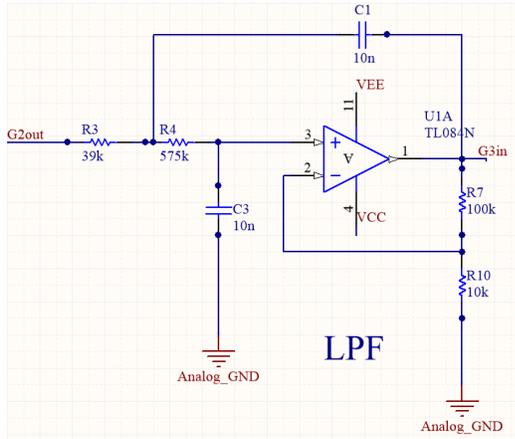
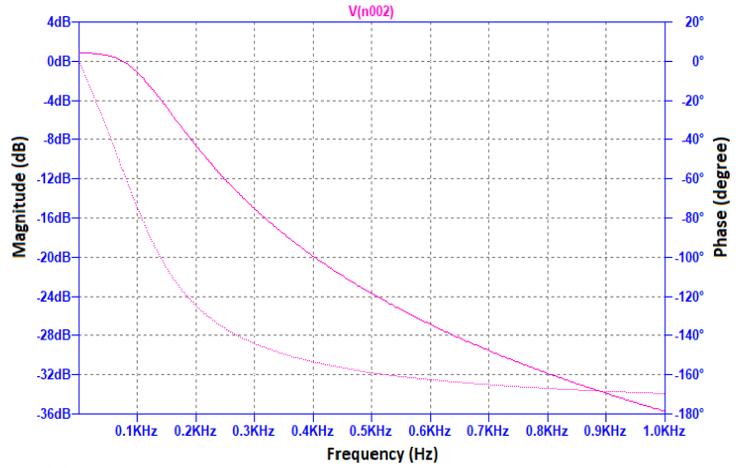

(a)

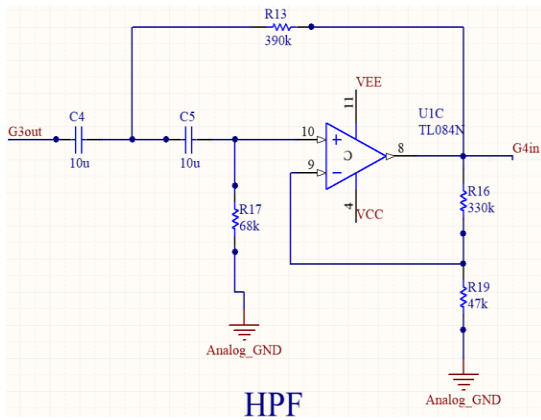
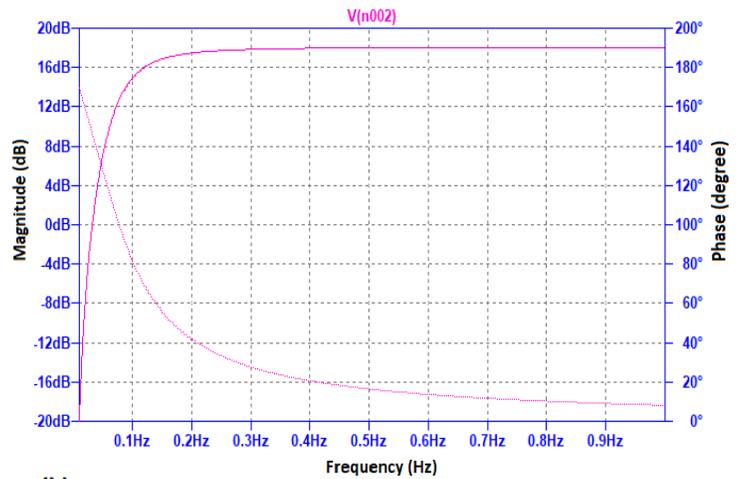

(b)

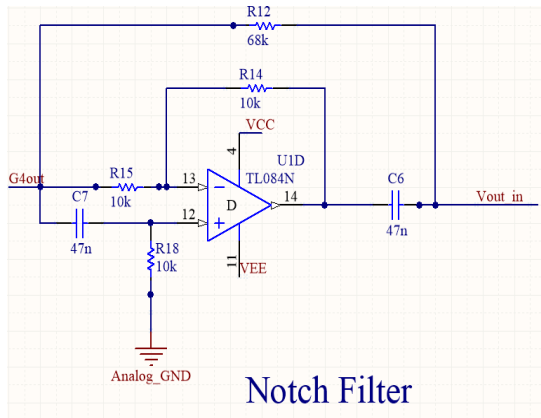
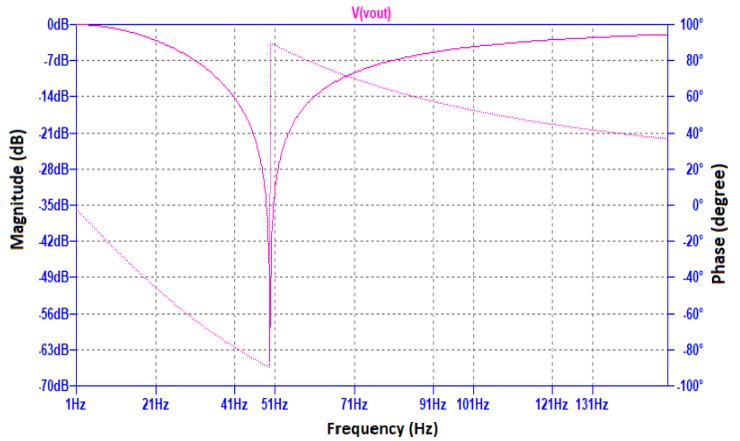

(c)

*Figure 3* Low-pass (a), high-pass (b), and notch filter (c) circuits along with their amplitude and frequency responses

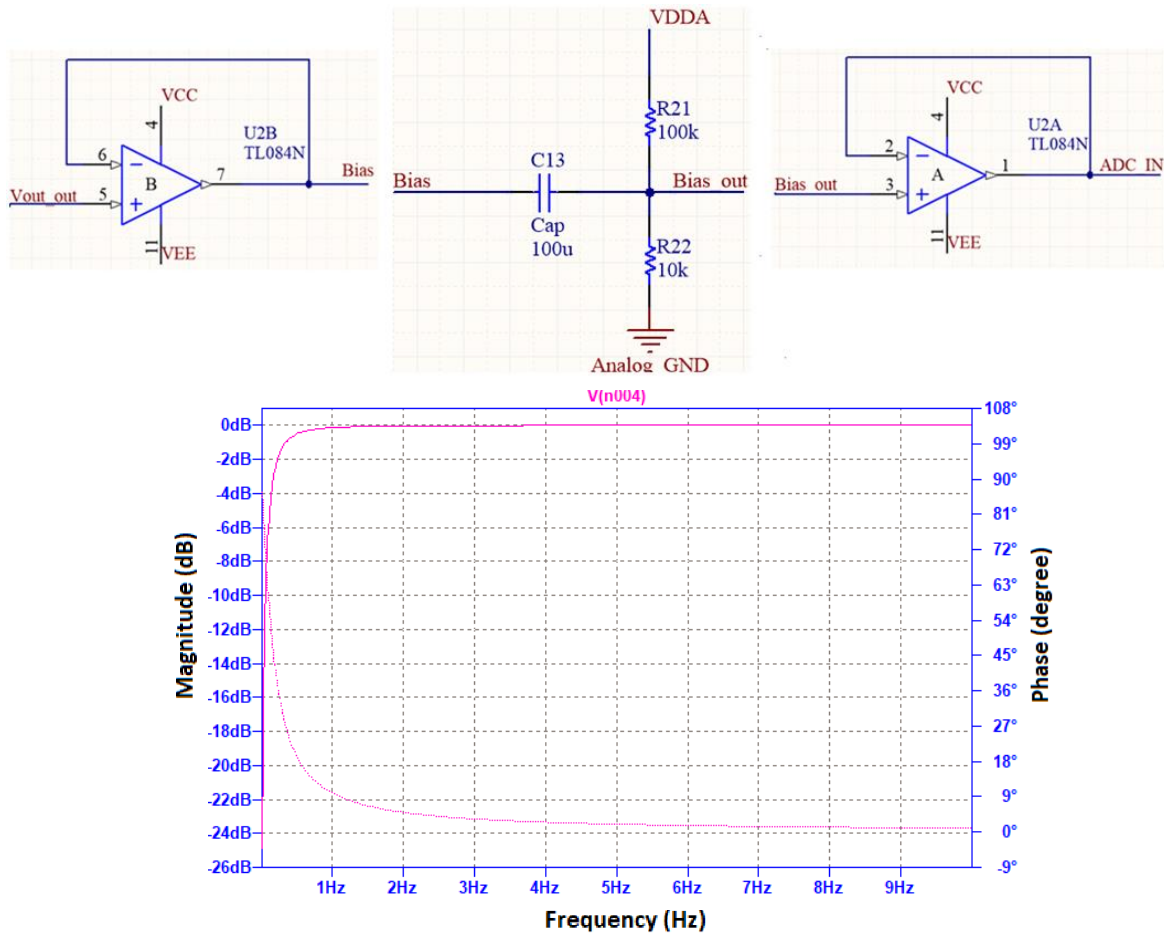

**Figure 4** *The circuit to bias the last stage output signal (top) and its frequency response (bottom)*

The reason for using buffers is to prevent the effect of voltage division loading on the last stage and ADC input. Furthermore, a relatively large 100uF capacitor is used to filter the DC components of the output signal. Figure 4 shows the frequency response of this part of the analog circuit.

### 2-2 Designation and implementation of the software

This section discusses the designation and implementation of the software needed to process the data received from the device's hardware.

#### 2-2-1 The STM32F030F4P6 microcontroller driving

As mentioned, the STM32F030F4P6 microcontroller is used to process data and transfer them to mobile. For driving this microcontroller, its clock frequency is set to 48 MHz, meaning each microcontroller's clock pulse will last 0.02083μs. The settings related to the clock frequency, ADC, timer, and USART are conducted in the STM32CUBE software.

#### 2-2-2 ADC setup and sampling using microcontroller's timer

The microcontroller's ADC resolution is 12 bits. According to its reference voltage, which is 3/3V, the minimum amount of voltage changes detected by the ADC is calculated using the equation 2 as the following formula:

$$Minimum\ Detectable\ Voltage\ Value = \frac{3.3}{2^{12} - 1} = 0.8059 mV \qquad (2)$$

Considering the gain we have in the analog part, this amount of resolution is enough to quantize the signal [17]. In the next step, we must determine at what rate the ADC should sample the signal data. According to the explanations given in the high-pass, low-pass, and notch filter sections about the available frequency range of the signal, the signal is sampled by a timer and a sampling frequency of 500/second. This sampling frequency is five times the signal's highest frequency component; Therefore, Nyquist's law regarding the sampling frequency has also been considered. For setting this sampling frequency using a timer, first, the ADC is set to the highest sampling rate, 3.55 Mbps, and employing the microcontroller's timer3 and setting its interrupt time to 2ms, with each interruption, the data is sampled from the ADC. Then the sampled data amplitude is converted to the voltage range from 0 to 3.3V and sent with the accuracy of 4 decimal places.

**2-2-3 Sending data to mobile using USART**

After receiving each sample and processing it in the interrupt body, the sample is sent to a mobile using the USART protocol. Considering that each data byte in this protocol must include a start and an end bit, each data byte is sent in the form of 10 bits. Also, since each signal sample has 4 decimal digits of accuracy, and a '\n' character must be sent after each sample to separate it from the others, each sample is sent in a 7-byte format. Therefore, with a sampling frequency of 500 samples/sec, the USART data transmission rate must be more than 10x7x500 bits per second so that there is no delay in sending data; therefore, the data transmission rate is set to 115200 bits/sec. We used HC-05 Bluetooth to send data wirelessly because its setup is straightforward, and its sending rate is adequate for sending signal data.

**2-2-4 Developing a mobile phone platform for displaying the electrocardiogram signal**

After sending signal data using Bluetooth, a mobile application is needed to receive these data to perform final post-processing and display them on the mobile screen. To create such a platform, we used Android Studio software, whose programming language is based on Java. In the following, the steps of designing this program are explained.

**2-2-4-1 Developing the front-end part of the platform**

The platform's front-end is a part of the developed platform that displays information to the user through which the user communicates with the back-end part of the platform. The designed front-end consists of two main tabs; when the program is started, the tab related to searching and connecting to Bluetooth is displayed first; then, in case of connection, the ECG signal is displayed to the user on the second tab. Figure 5 shows the designed front-end.

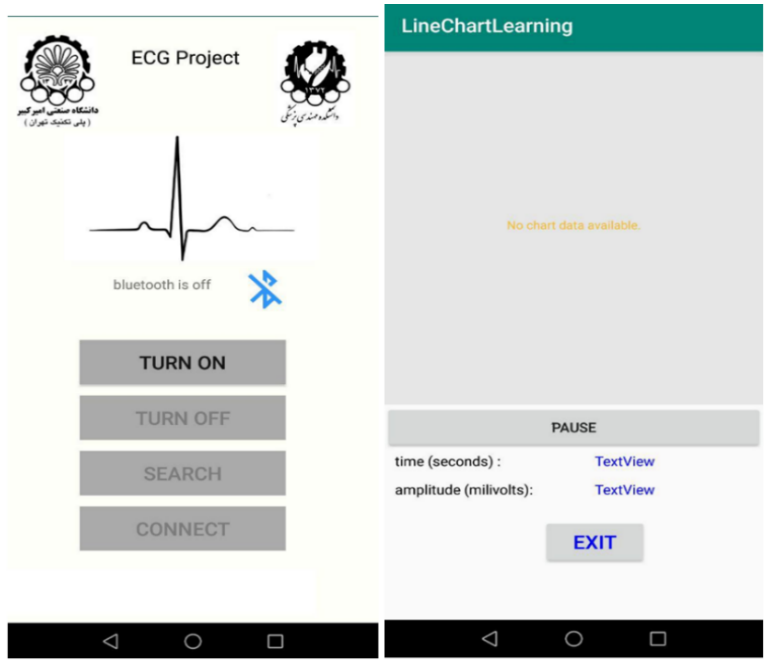

*Figure 5 The designated front-end of the mobile platform*

**2-2-4-2 Developing the back-end part of the platform**

This part is related to the coding part of the platform for receiving and processing data; without this part, the program's front-end has practically no usage. The back-end design of this project includes three parts: searching, connecting, and receiving Bluetooth data, processing data, and displaying them using the lineChart class.

**2-2-5 Digital filters**

For the final data filtering, an FIR bandpass filter with an order of 500 and a notch IIR filter with an order 6 are used. We used the FIR filter of type least square and the IIR filter of type Butterworth. The low cut-off frequency of the bandpass filter is 1Hz and the high cut-off frequency of this filter is 102Hz; this way, the baseline wander, the power line noise, and the high-frequency noise of the signal are filtered. The memory required to implement these two filters is 11KB; therefore, due to the microcontroller's limited memory, the filters were implemented in the mobile-written platform. Figures 6 and 7 show the frequency response, phase response and group delay related to notch and bandpass filters.

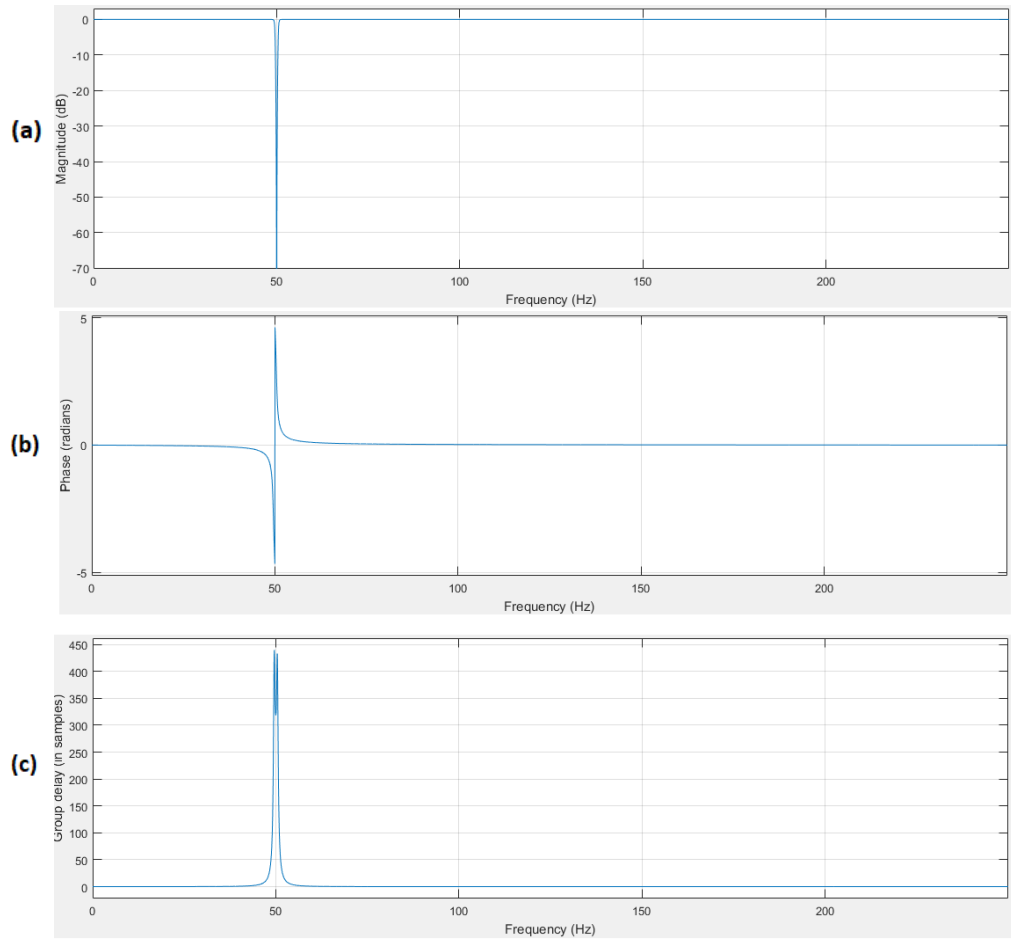

*Figure 6* The amplitude (a), phase response (b), and group delay (c) of the implemented IIR digital filter in mobile platform

### 2-2-6 The value of delay in the output signal

FIR filters cause a delay in the output signal, in which the amount of the delay in seconds is calculated using the following formula:

$$d = \frac{(FL - 1)}{2 \times SR} \quad (3)$$

In which SR is the sampling rate in Hz and FL is the length of the used filter [15]. Considering that the length of the used digital filters is 500 and the sampling rate is 500Hz, the output signal delay is approximately 0.5s. Figure 8 shows the ECG signal in two modes. The right and left figures correspond to the filtered and unfiltered signal modes.

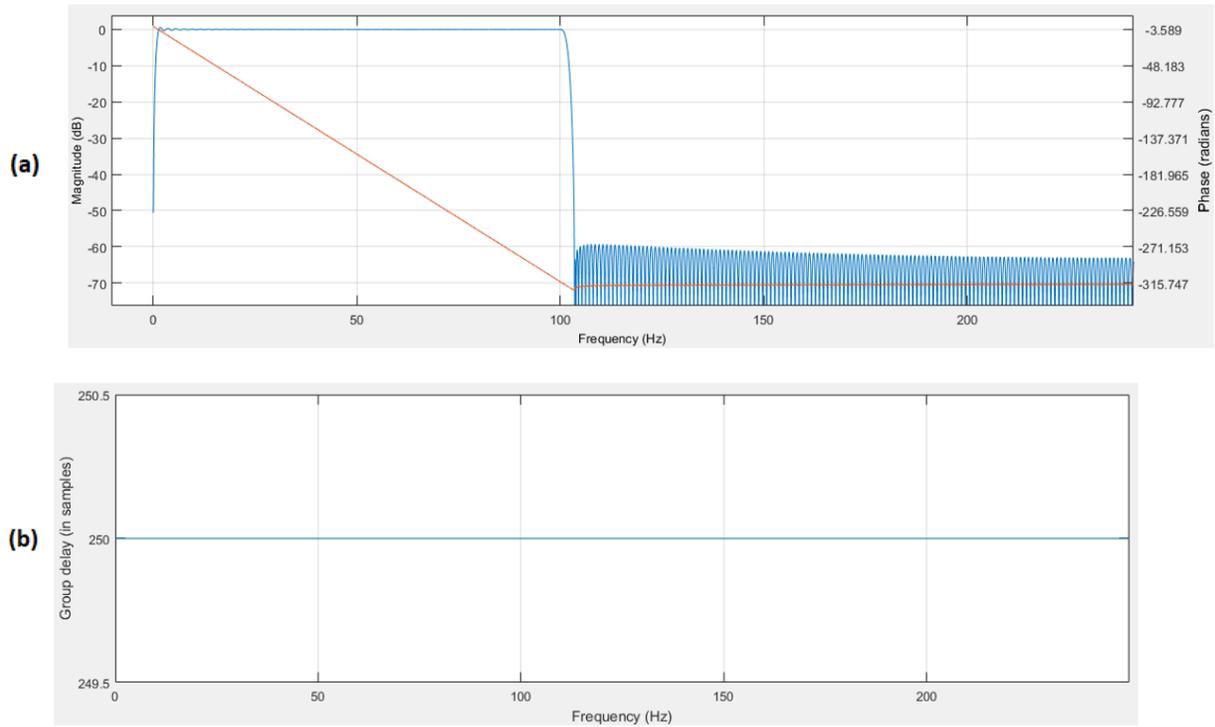

*Figure 7* The amplitude (a) and phase response (b) and group delay of the implemented FIR digital filter in mobile platform

### 2-2-7 Beat rate extraction algorithm

To compute the number of heartbeats, we usually measure the distance between two R waves and calculate the number of beats per minute according to the sampling rate. The number of heart beats per minute is calculated as follows:

$$BPM = \frac{60 \times SR}{(n_{i+1} - n_i)} \qquad (4)$$

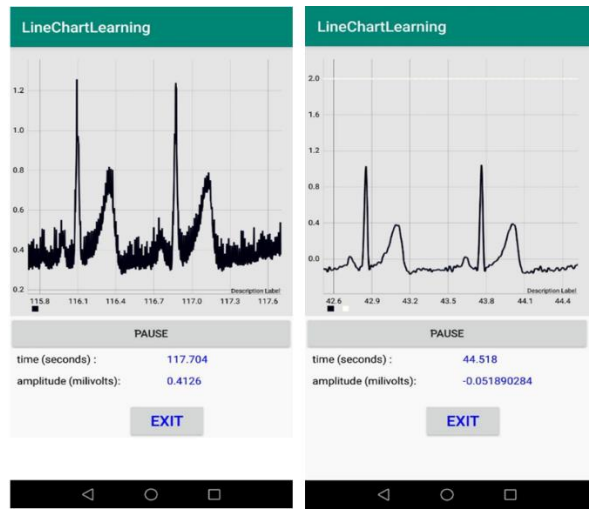

*Figure 8* The signal displayed on the mobile screen in two modes, unfiltered (left) and filtered (right) modes

Where $n_i$ and $n_{i+1}$ are $i^{th}$ and $i+1^{th}$ samples taken from the ECG and SR signals, respectively. There are different methods to find the occurrence time of R waves. In this project, we used the most basic pulse rate extraction algorithm written by Pan Tomkins, in which the amount of error in the calculation is minimized [18]. This algorithm calculates the pulse rate in two ways; firstly, after performing preprocessing on the signal and filtering the interfering signals using identified R-waves; secondly, using the algorithm shown in Figure 9, the beat rate is calculated. Finally, the results obtained from both approaches are compared with each other, and if the R wave is correctly identified in both cases, an R wave is reported in the output; otherwise the R wave is rejected [18,19]. Each of these approaches cannot be used alone to identify the beat rate because in the algorithm used in Figure 9, if the noise amplitude is high, the derivative block can amplify it and cause an error in the identified R waves. Also, in some patients, due to the high amplitude of the P wave, there is a possibility that this wave will be recognized as an R wave. As a result, it is not possible to use only the main signal to calculate the heart rate [18,19]. Combining the results of these two approaches will reduce the percentage of the abovementioned errors. A point that must be considered in comparing the results of these two approaches is the value of signal delay in each of them; this way, by calculating the delay that digital processing creates in each of the approaches, the signals must also be equally time shifted, and the results should be compared with each other [18,19]. Since in both approaches, the signal first passes through the notch and bandpass filters, the delay in both of them is equal after pre-processing. In the second case, the signal passes through a derivative block in which its memory is 13 samples; a 26ms delay is added to the previous delays. The square block has no delay; therefore, the difference between the first and second approaches is only 26ms meaning that the first state must be shifted by 26ms in time. After the explained time shift, the results related to the first and second approaches can be aligned.

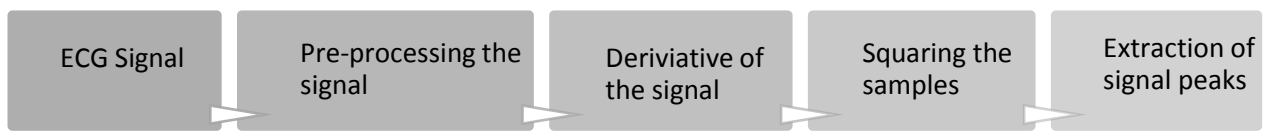

*Figure 9* Block diagram of the method of detecting ECG signal peaks and calculating the heart rate [18,19]

Since the first step of the algorithm shown in Figure 9 has already been done, the derivative of the ECG signal is first calculated and then squared. The ECG signal, its derivative, and the squared data of the derivative signal are displayed in the mobile application as Figure 10. Finally, thresholding is used to distinguish the peaks and their distance from each other; after comparing the output results of the two different approaches to heart rate calculation in the algorithm written by Pan-Tomkins; if the results of both modes are correct, the time index corresponding to the detected R peak is stored in a local variable. The BPM rate is calculated by detecting the next peak and using equation 4.

## 3 Results and Discussion

This section discusses the performance of the designed device and the achieved results. First, the referred noise to the input and the ECG signal noise, as well as their power spectral density, are measured to check the device's performance. Then, the ECG signal of 10 subjects was acquired for a certain period of time to calculate the accuracy and error percentage of the algorithm implemented to calculate the heart rate.

### 3-1 Output signal noise measurement

The signal and noise power spectral densities and the signal-to-noise ratio are calculated to check the quality of the ECG signal recorded by the device. For computing the power spectral density, we must first calculate the signal FFT; therefore, we can use the Noise class in Android Studio. Note that FFT is the DFT whose implementation has been optimized [17]. The DFT formula is given as the following relation:

$$X[k] = \sum_{n=0}^{N-1} x[n]e^{-jk\frac{2\pi}{N}n} \qquad 0 < k < N-1 \qquad (5)$$

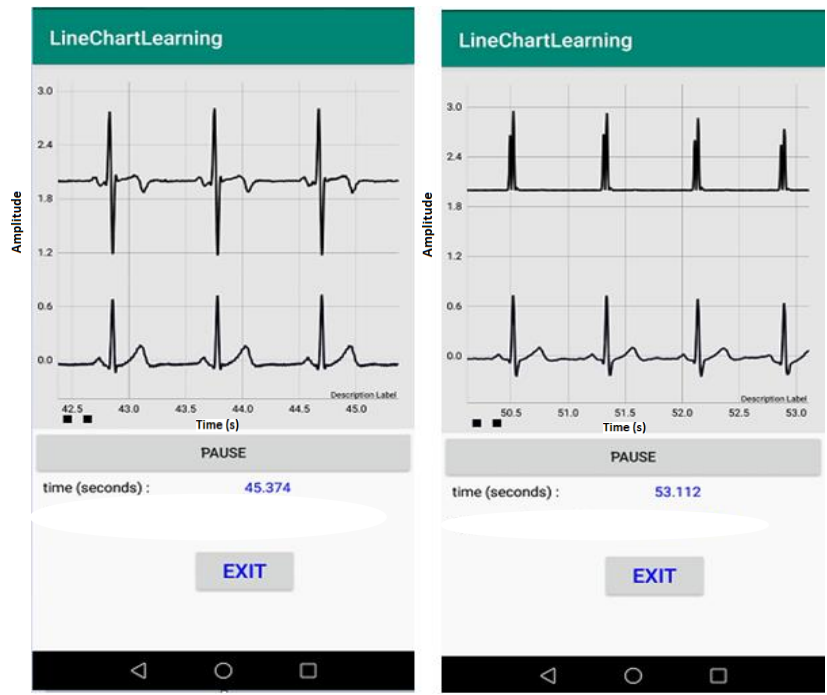

**Figure 10** *Display of the ECG signal, its derivative (left) and the square of the signal derivative (right) in the mobile platform (note that the derivative of the signal is shifted to the right by 26ms in time compared to the signal itself)*

Where X[k] is the Fourier transform, x[n] is the discrete time signal, and N is the number of points used to calculate the FFT. According to the above relationship, the bigger N is, the more frequency resolution we will have. Hence, 4096 points are used in the FFT calculation for sufficient resolution in the frequency domain. After calculating the FFT, the power spectral density of the signal and noise is calculated using the following formula:

$$PSD = 10\ log10(|X[k]|^2/(N \times SR)) \tag{6}$$

Where $|X[k]|^2$ is the square of the signal FFT, and SR is the sampling rate. Figure 11 shows the power spectral density of the noise referred to the input and the ECG signal acquired from Lead I; in which the signal-to-noise ratio is about 50dB.

The sudden decrease in the power spectral density in the Figure 11 is the stop band used in the low-pass filter, which filters the frequencies above 102Hz. Figure 12 shows the ECG signals acquired from Leads I and II, each of which has a different gain. It can be seen that the signal amplitude is stronger in Lead II, and the quality of the acquired signal is also better. The value of the gains in Lead I are equal to 600 and 1800, and in Lead II are equal to 250 and 300.

## 3-2 The accuracy of the heart rate calculation algorithm

To investigate the accuracy and error of the written algorithm in the calculation of heart beat rate, the ECG signals of 10 subjects were recorded for 60s. Then, different thresholds were applied to the signal to detect the signal peaks. According to Figure 13, the number of detected peaks in the signals varies using different thresholds.

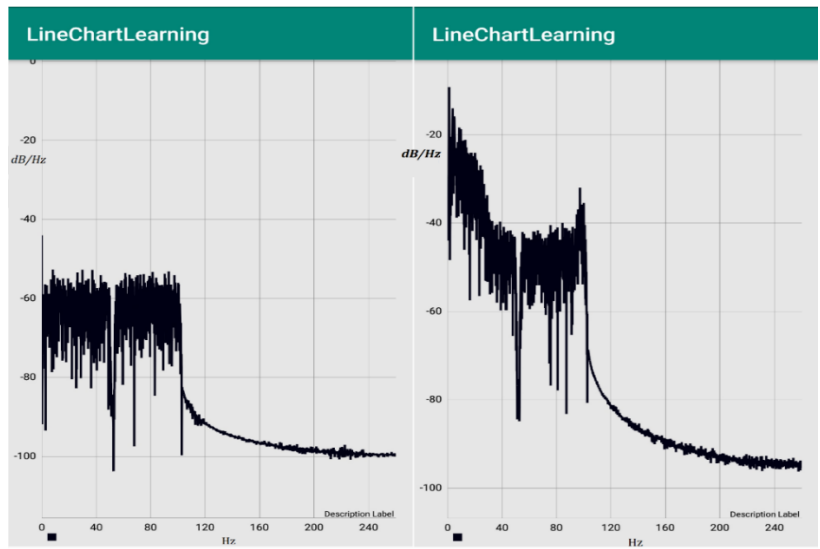

*Figure 11* The noise power spectral density referred to the input (left) and the resulting ECG signal power spectral density from lead I (right)

According to the figure above, when the threshold value is low, many local maxima, in addition to the main peaks, are wrongly detected; increasing the threshold value decreases the number of found maxima such that after a specific value, the main peaks are also missed. The receiver operating characteristic (ROC) curve of the algorithm written for determining the heart rate of one subject is shown in Figure 14. To choose the optimal threshold value, the ECG signals of 10 persons were recorded, and their ROC curve was calculated; then, the thresholds, shown with a red arrow in the ROC curve, in which the true positive rate (TPR) and true negative rate (TNR) criteria were 100% were averaged and the obtained value was considered as the final threshold.

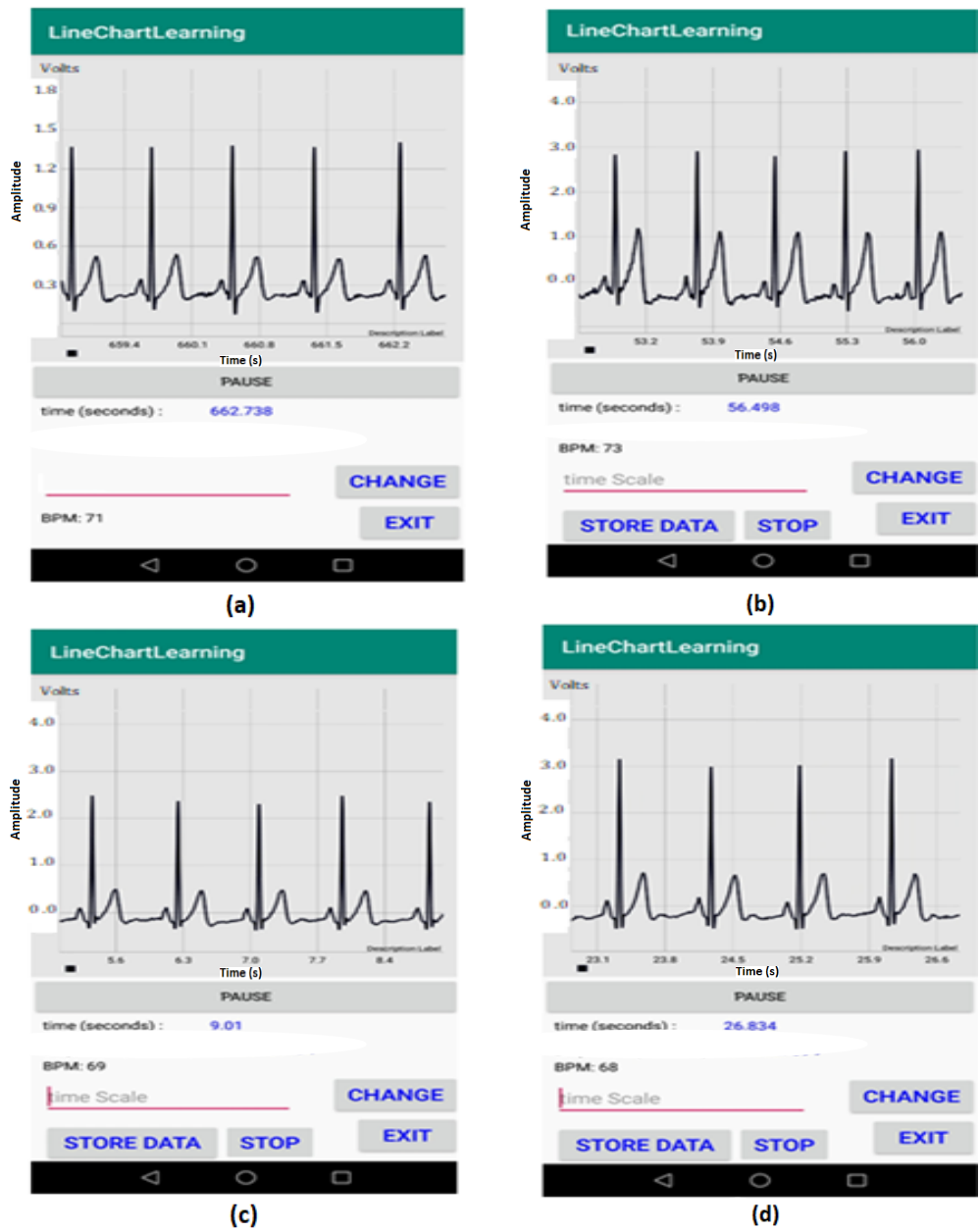

*Figure 12* The recorded ECG signal from Lead I (a,b) and Lead II (c,d)

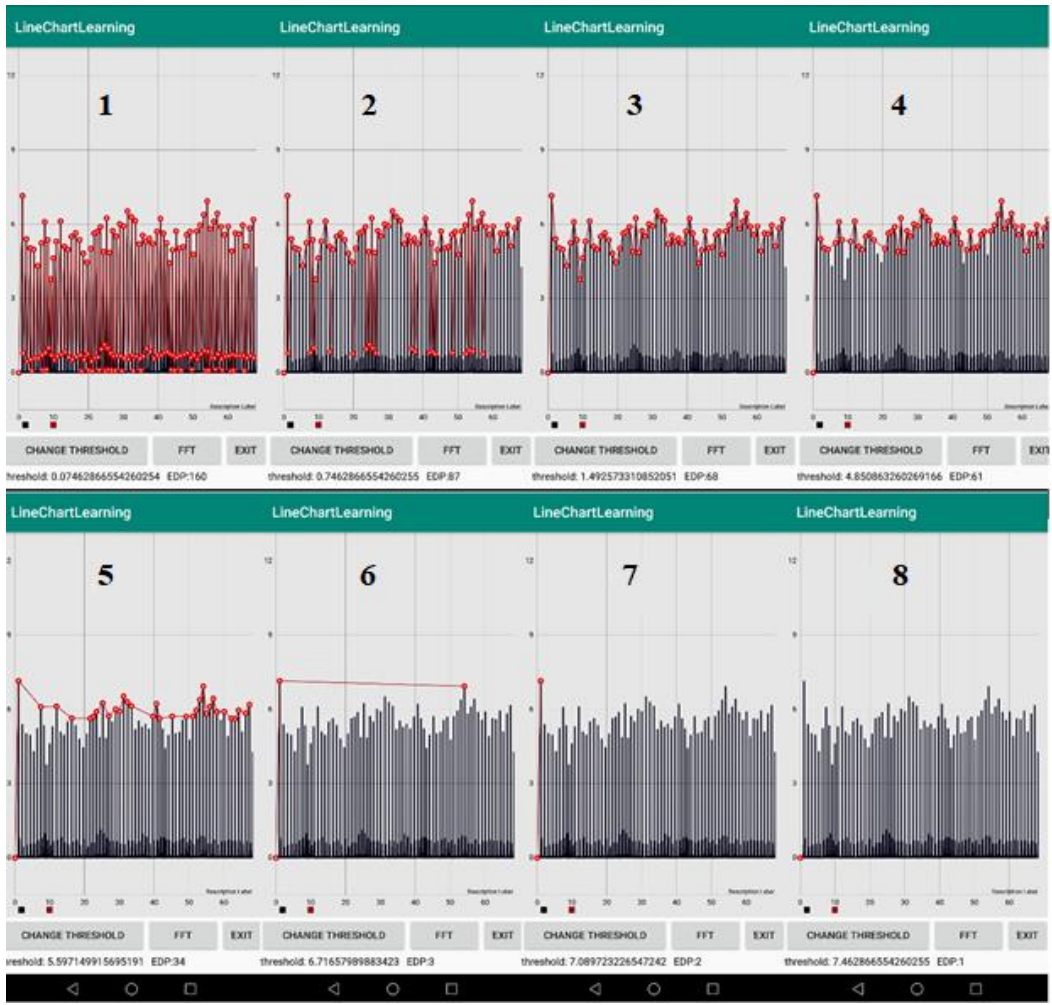

*Figure 13* ECG signal recording and thresholding to detect the signal peaks (it can be seen that in steps 1 and 2, a very low threshold was used to detect the peaks; therefore, many points in addition to the valid points have been detected. By increasing the threshold value in step 3, the number of incorrectly detected points vanishes. From steps 4 to 7, the number of correctly detected peaks decreases with the threshold increase. Finally, the number of the detected points reaches zero in the 8th step)

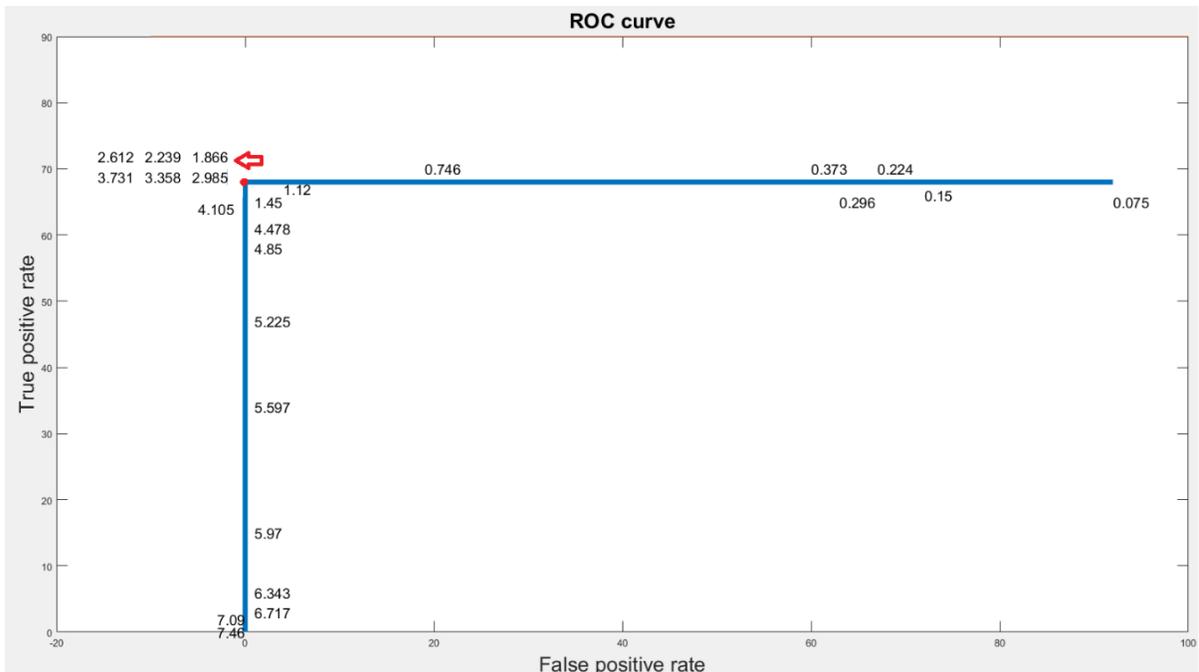

*Figure 14* ROC curve related to different thresholding to detect signal peaks

## 4 Conclusion

In this project, we aimed to design an ECG recording device to simplify healthcare of patients so that the necessity to a nurse or doctor to install the electrodes on body be removed. Thus, we fabricated a device to record the ECG signal using a single Lead. The signal is first amplified and filtered by an analog circuit; then, using a digital circuit, it is sampled and transmitted to a mobile phone through the USATR protocol and using the Bluetooth module. We employed AD620 and TL084 amplifier ICs in the analog circuit and STM32F030F4P6 microcontroller in the digital circuit. In the power supply section of the device, LM1117 was used to supply the microcontroller and transmitter, and TC7660 was used to supply the negative voltages of the circuit. Finally, we developed a platform using Android Studio to display the recorded signal on the mobile phone. The platform receives, processes, and displays the signal data from the transmitter. Furthermore, to compute the heart beat rate, the algorithm written by Pan Tomkins was employed. Eventually, to check the quality of the output ECG signal, the power spectral density of the signal and the noise referred to the input were calculated and plotted. The result was promising, and we achieved a signal-to-noise ratio of 50dB.

## Supplementary material

The picture of the fabricated device is illustrated in Figure 15. Also, a video demonstration of the ECG signal acquisition on the mobile screen can be found at:

https://drive.google.com/file/d/1Wmv8YtMftGVUnlr0LrX8gYYeRRrKpqN3/view?usp=sharing

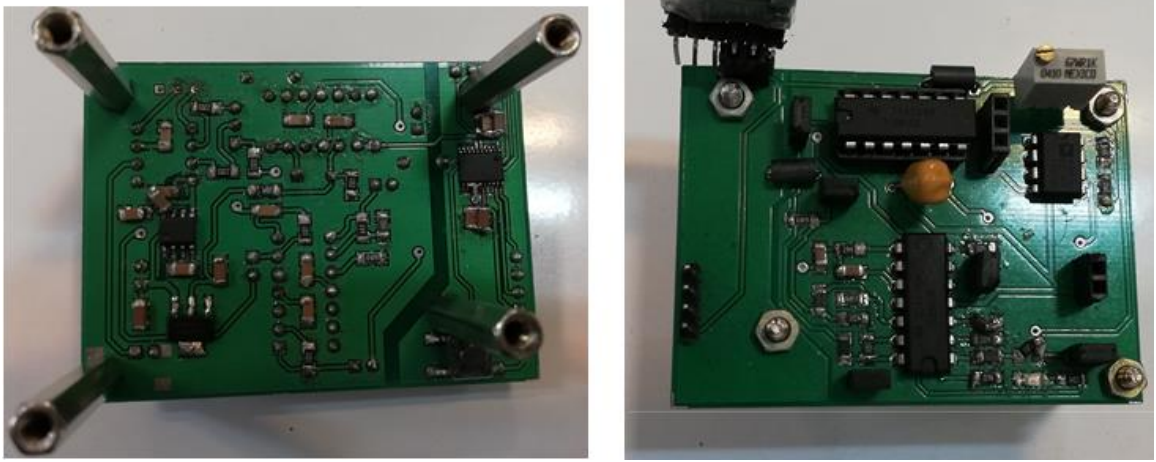

*Figure 15 The fabricated ECG signal acquisition device*